\documentclass[prl,preprintnumbers,twocolumn,amsmath,amssymb]{revtex4}%twocolumn preprint
\usepackage{cases}
\usepackage{txfonts}
\usepackage{amssymb}
\usepackage{amsmath}
\usepackage{graphicx}
\usepackage{wrapfig}
\usepackage{dcolumn}
\usepackage{bm}
\usepackage{wasysym}

\begin{document}

\title{Effects of variable anisotropic-strain on the emission of neutral excitons confined in epitaxial quantum dots}

\author{J. D. Plumhof$^{1*}$\footnote[0]{$^{\text{*)}}$Electronic mail: j.d.plumhof@ifw-dresden.de}}
\author{V. K\v{r}\'apek$^{2}$}
\author{ F. Ding$^{1,3}$}
\author{K. D. J\"ons$^{4}$, R. Hafenbrak$^4$}
\author{P. Klenovsk\'y$^{2}$}
\author{A. Herklotz$^{5}$, K. D\"orr$^5$}
\author{P. Michler$^{4}$}
\author{A. Rastelli$^{1\dag}$\footnote[0]{$^{\text{\dag)}}$Electronic mail: a.rastelli@ifw-dresden.de}}
\author{ O. G. Schmidt$^1$}

\affiliation{$^1$ Institute for Integrative Nanosciences, IFW
Dresden, Helmholtzstr. 20, D-01069 Dresden, Germany}
\affiliation{$^2$ Institute of Condensed Matter Physics, Masaryk
University, Kotl\'a\v{r}sk\'a 2, 61137 Brno, Czech Republic}
\affiliation{$^3$Key Laboratory of Semiconductor Materials Science,
Institute of Semiconductors, Chinese Academy of Sciences, Beijing
100083, China} \affiliation{$^4$ Institut f\"ur Halbleiteroptik und
Funktionelle Grenzfl\"achen, University of Stuttgart, Allmandring 3,
70569 Stuttgart, Germany} \affiliation{$^5$ Institute for Metallic
Materials, IFW Dresden, Helmholtzstr. 20, D-01069 Dresden, Germany}

\date{\today}

\begin{abstract}
We study the effect of elastic anisotropic biaxial strain on the
light emitted by neutral excitons confined in different kinds of
semiconductor quantum dots (QDs). We find that the light
polarization rotates by up to $\sim$80$^\circ$ and the excitonic
fine structure splitting varies by several tens of $\mu$eVs as the
strain is varied. By means of a continuum model we mainly ascribe
the observed effects to substantial changes of the hole wave
function. These results show that strain-fields of a few $\permil$
magnitude are sufficient to dramatically modify the electronic
structure of QDs.

\end{abstract}

\maketitle

Semiconductor quantum dots (QDs) obtained by epitaxial growth are
receiving much attention because of their potential use as building
blocks for quantum information processing and communication
devices~\cite{Zrenner10,Press08,Brunner09,Akopian06,Hudson07,Hafenbrak07,Dousse10}.
QDs confine the motion of charge carriers in three-dimensions and
are thus referred to as artificial atoms. Similar to real atoms,
external electric and magnetic fields can be used to manipulate the
properties of bound states in
QDs~\cite{Empedocles97,Bayer02,Gerardot07,Vogel07,Stevenson06}.
 In addition, the solid-state character of QDs allows for engineering methods
which are not available for atoms. Dynamic stress
fields~\cite{Itskevich98,Seidl06,Ding10,Metcalfe10} represent an
example, whose wide potential is only recently being
recognized~\cite{Bryant10, Singh10}.

The structural properties of QDs (crystal structure, material
composition, shape and size) mainly determine their electronic and
optical properties. In particular, the emission of neutral excitons
confined in QDs with symmetry lower than $D_{2d}$ is typically split
by several tens of $\mu$eV because of the anisotropic electron-hole
exchange interaction~\cite{Bayer02,Gammon96,Ramirez2010}. This
broken degeneracy of the bright excitonic states, referred to as
fine structure splitting (FSS) prevents the use of QDs as sources of
entangled photon pairs on
demand~\cite{Benson00,Akopian06,Hudson07,Hafenbrak07,Dousse10}.
External electric or magnetic fields have been applied to restore
the QD symmetry and achieve FSS values comparable to the radiative
linewidth~\cite{Gerardot07,Vogel07,Itskevich98}. Seidl \emph{et
al.}~\cite{Seidl06} showed that also uniaxial strain can in
principle be used to reduce the excitonic FSS. Due to the limited
tuning range available, it has however remained unclear whether
strain is suitable to reach sufficiently low values of
FSS~\cite{Hudson07}, and what the mechanisms behind the observed FSS
changes are.

Recently it was predicted, based on atomistic model simulations for
InGaAs/GaAs QDs~\cite{Singh10}, that anisotropic stress acting on
QDs leads to an anticrossing of the bright excitonic states.
Thereby, the magnitude and phase of the mixing of the bright
excitonic states are modified which results in a change of the FSS
and in a rotation of the linear polarization of the emitted
photons~\cite{Bryant10}. Here we present a detailed experimental
proof of this effect and explain its origin by using a continuum
model based on 8-band k$\cdot$p and configuration interaction
theory.

The measurements are performed on two different samples grown  by
solid-source molecular beam epitaxy (MBE). The active structures
consist of QDs embedded in thin membranes, which are released from
the underlying substrate and transferred onto a piezoelectric
actuator~\cite{Zander09,Ding10}. The first membrane sample, with
total thickness of about 150~nm, contains GaAs/AlGaAs
QDs~\cite{Plumhof10} and quantum well (QW) potential fluctuations
(QWPFs)~\cite{Gammon96,Plumhof10}. The latter, which are produced by
local thickness or alloy fluctuations in a narrow QW, confine the
carriers in all three dimensions and act as QDs with low confinement
potential. The second sample contains standard InGaAs/GaAs QDs
embedded in 200~nm thick membranes~\cite{Supplementary}).

\begin{figure}[h]
\centering
\includegraphics[width =0.4 \textwidth]{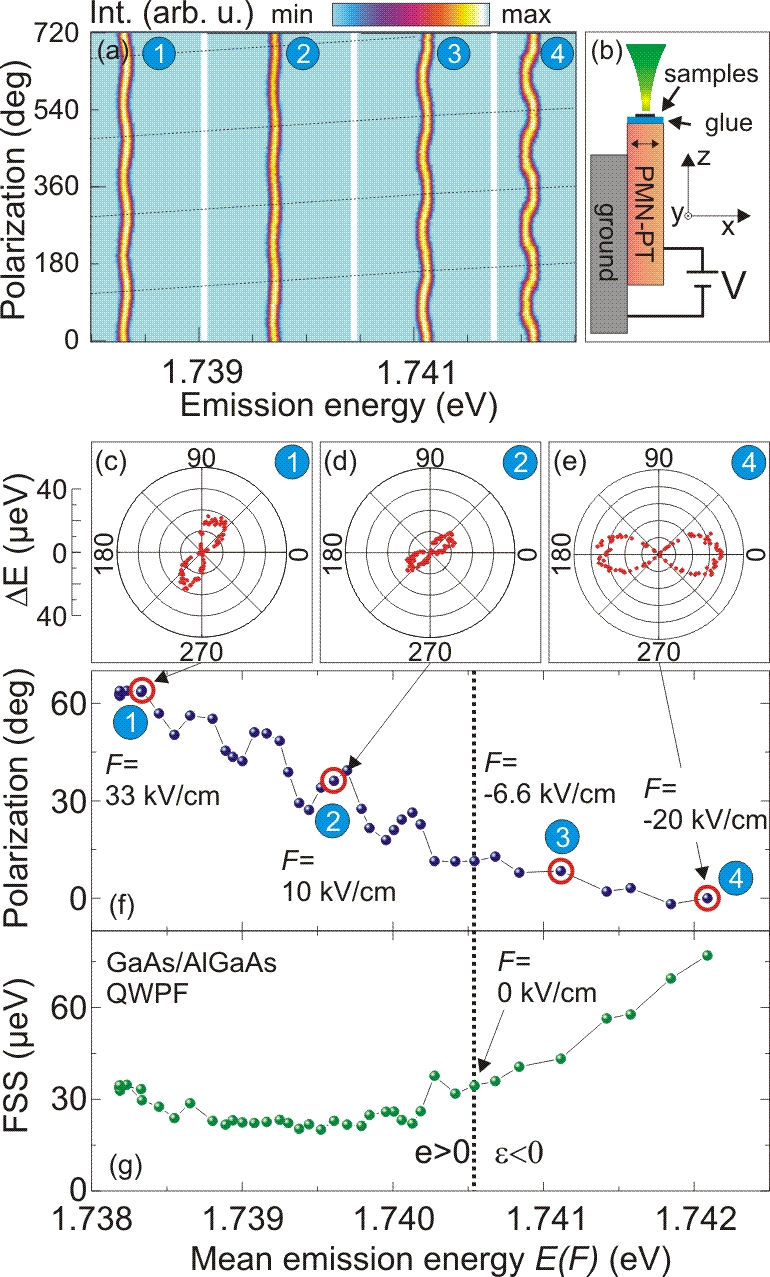}\caption{
(Color online) Behavior of a neutral exciton confined in a
GaAs/AlGaAs QWPF under anisotropic biaxial stress. (a) Color-coded
PL intensity vs polarization angle and energy for different values
of electric fields applied to the piezoelectric actuator  [the field
values for the panels 1 to 4 are indicated in (f)]. The dashed lines
are guides to the eye showing the rotation of the polarization
direction. The x direction in (b) corresponds to polarization angles
of 0, 180, 360, 540 and 720 degrees and coincides with the
polarization direction of the high-energy component at $F=-$20~kV/cm
(panel 4). (b) Sketch of the device consisting of a membrane
(sample) glued on a side of a PMN-PT crystal. (c, d, e) Polarization
dependence in polar coordinates of the relative emission energy for
panels 1, 2 and 4 of (a). The mean emission energy $E(F)$ for each
value of $F$ is subtracted. (f) Polarization angle of the
high-energy component of X with respect to the direction of the
electric field vs $E(F)$. The dots marked by red circles correspond
to the data shown in (a). (g) FSS vs average emission energy.}
\label{fig:rawdata}
\end{figure}

The external stress is applied using a piezoelectric
$\mathrm{[Pb(Mg_{1/3}Nb_{2/3}O_3]_{0.72}-[PbTiO_3]_{0.28}}$ (PMN-PT)
crystal. By applying a voltage $V$ between the front and the back
surface of the crystal [i.e. along the $x$ axis in
Fig.~\ref{fig:rawdata}(b)] the side faces, such as the top $x-y$
plane, expand (or contract) parallel to the direction of the
electric field $F$, for positive (negative) applied voltage.
Simultaneously, the side faces contract (or expand) perpendicular to
the electric field [i.e. along the $y$ axis in
Fig.~\ref{fig:rawdata}(b)]. We denote the strain parallel to the $x$
and $y$ axes as $\varepsilon$ and $\varepsilon_{\bot}$,
respectively. The relation between these strain components is
$\varepsilon_{\bot}\approx -0.7 \times \varepsilon$ (see
Ref.~\onlinecite{Biegalski2010}).
 By placing membranes with QDs on the side faces of the
PMN-PT we can thus apply strongly anisotropic biaxial stress on the
QDs. According to previous results~\cite{Zander09,Ding10}, we expect
values of $\varepsilon$ of the order of a few $\permil$ in the
explored range of the electric field $F$.

%
% Added "linear"
%
Photoluminescence (PL) spectroscopy measurements are performed at a
temperature of 8~K in a standard micro-PL setup with a spectral
resolution of about 70~$\mu$eV. The linear polarization of the
luminescence is analyzed by combining a rotatable achromatic
half-wave plate and a fixed linear polarizer~\cite{Supplementary}.

Figure~\ref{fig:rawdata}(a) shows a color-coded PL-intensity map for
a neutral exciton (X) confined in a GaAs/AlGaAs QWPF as a function
of the emission energy and polarization angle for different values
of the electric field $F$ applied to the PMN-PT (panels 1 to 4:
$F$=33, 10, $-$6.6 and $-$20~kV/cm). In this membrane the electric
field direction forms an angle of about 20$^{\circ}$ with
 the [1$\overline{1}$0] GaAs crystal direction. The polarization dependent
 periodic energy shift (wavy pattern)
 observed in PL is ascribed to the excitonic FSS (see
Ref.~\onlinecite{Plumhof10}).

Two striking features clearly emerge from Fig.~\ref{fig:rawdata}(a):
(i) The polarization direction of the excitonic emission, related to
the phase of the wavy pattern, rotates by more than 60$^\circ$ when
$F$ is changed from 33 to $-$20~kV/cm (see dotted lines); (ii) The
magnitude of the FSS, i.e. the amplitude of the oscillations of the
wavy patterns, first decreases and then increases with decreasing
electric field. To extract quantitative information from the data,
we first fit the peak position with a single Lorentzian curve at
each polarization angle. The obtained relation of peak position vs
polarization angle is then fitted by a sine function to estimate
both the magnitude of the FSS and the polarization of the X
transitions with respect to the field direction [$x$ axis in
Fig.~\ref{fig:rawdata}(b)]. The resolution in the determination of
the FSS with this procedure is around 2.5~$\mu$eV~\cite{DingBohm10}.

%
% Slightly revised Fig caption - added "relative"
%

%
% Revised slightly description of (f) to make clear what we see.
% Always tell the reader what he should note - the oscillations were not mentioned at all
%
Figure~\ref{fig:rawdata}(c)-(e) shows polar plots of the relative
peak positions $\Delta$E extracted from Fig.~\ref{fig:rawdata}(a)
after subtraction of the average emission energy $E(F)$ measured for
different values of $F$. The data, which are averaged over two
periods of the polarization-resolved measurements (from 0$^\circ$ to
360$^\circ$ and from 360$^\circ$ to 720$^\circ$),  clearly show the
strain-induced changes both in polarization direction and FSS.
Figure~\ref{fig:rawdata}(f) shows the orientation of the linear
polarization of the high energy component of X (with respect to the
direction of $F$) as a function of $E(F)$.
Figure~\ref{fig:rawdata}(g) shows the corresponding behavior of the
FSS. When moving from low to high emission energies, i.e. from
tensile to compressive strain along x, the FSS goes through a broad
minimum before increasing again. The maximum observed change of the
FSS for this QWPF is about 50~$\mu$eV. Concerning the polarization
direction we observe that: (i) It shows oscillations superimposed to
a smooth decrease when the FSS is minimum; (ii) It
 appears to saturate with increasing FSS and, more precisely, it is
 aligned parallel to $F$ for strong compression (point 4).
By repeating similar measurements on different QWPFs we consistently
observe the same qualitative behavior: the polarization rotation
mainly occurs in correspondence to the minimum of the FSS.
Interestingly, the minimum value of the FSS varies from one QWPF to
another. Examples where the minimum FSS reaches values below about
5~$\mu$eV are presented in Fig.~\ref{fig:QDs}(d) and in the
supplementary information~\cite{Supplementary}.

In order to test whether the effects observed for GaAs/AlGaAs QWPFs
occur also for other QD structures, we have performed similar
experiments with GaAs/AlGaAs QDs, and InGaAs/GaAs QDs
(see Fig.~\ref{fig:QDs}).
\begin{figure}[h]
\centering
\includegraphics[width =0.45 \textwidth]{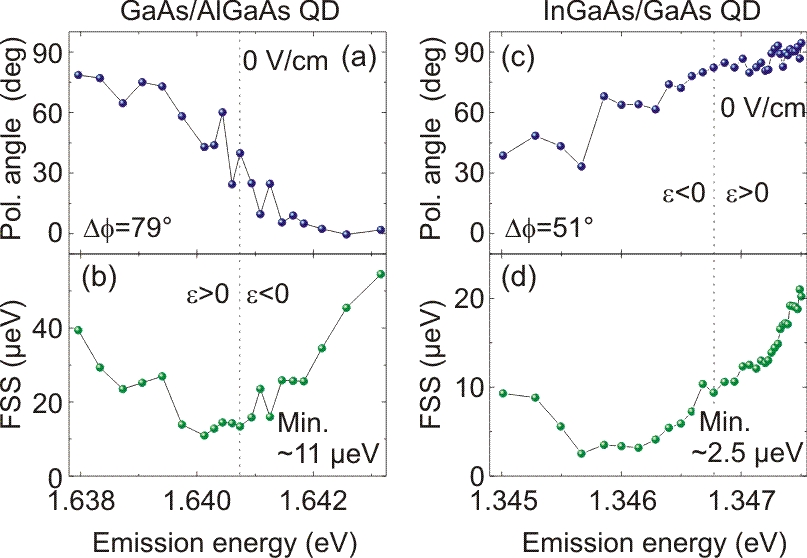}\caption{
(Color online) (a,b) Polarization and FSS behavior of a GaAs/AlGaAs
QD and (c,d) of an InGaAs/GaAs QD. The polarization angle $\phi$ is
defined as the angle of the higher energetic emission line with
respect to the x direction in Fig.~\ref{fig:rawdata}(b). For both
QDs shown here the x direction roughly corresponds to the
[1$\overline{1}$0] direction of the GaAs-membrane.}
\label{fig:QDs}\end{figure} For compressive strain, the emission
energy of the GaAs/AlGaAs QD presented in Fig.~\ref{fig:QDs}(a) and
(b) blue shifts similar to the QWPFs, whereas the emission energy of
the presented InGaAs/GaAs QD [Fig.~\ref{fig:QDs} (c) and (d)] red
shifts. In spite of the very different structural properties and
behavior of the emission energy under anisotropic strain, we find
that the excitonic behavior of the two different types of QDs is
qualitatively the same as that observed for the GaAs/AlGaAs QWPFs:
we observe a clear polarization rotation of the emitted light (by
79$^{\circ}$ for the GaAs/AlGaAs QD and by 51$^{\circ}$ for the
InGaAs/GaAs QD) [see Fig.~\ref{fig:QDs} (a) and (c)].
Simultaneously, the FSS is tuned in a range of $\sim$~70~$\mu$eV for
the GaAs QD and $\sim$~25~$\mu$eV for the InGaAs QD [see
Fig.~\ref{fig:QDs} (b) and (d)]. Furthermore, we observe that the
rotation of the polarization mainly takes place when the FSS reaches
its minimum value. As in the case of the QWPF presented in
Fig.~\ref{fig:rawdata}, significant fluctuations of the polarization
angle are observed in correspondence to the minimum of the FSS.

\begin{figure}[h]
\centering
\includegraphics[width =0.45 \textwidth]{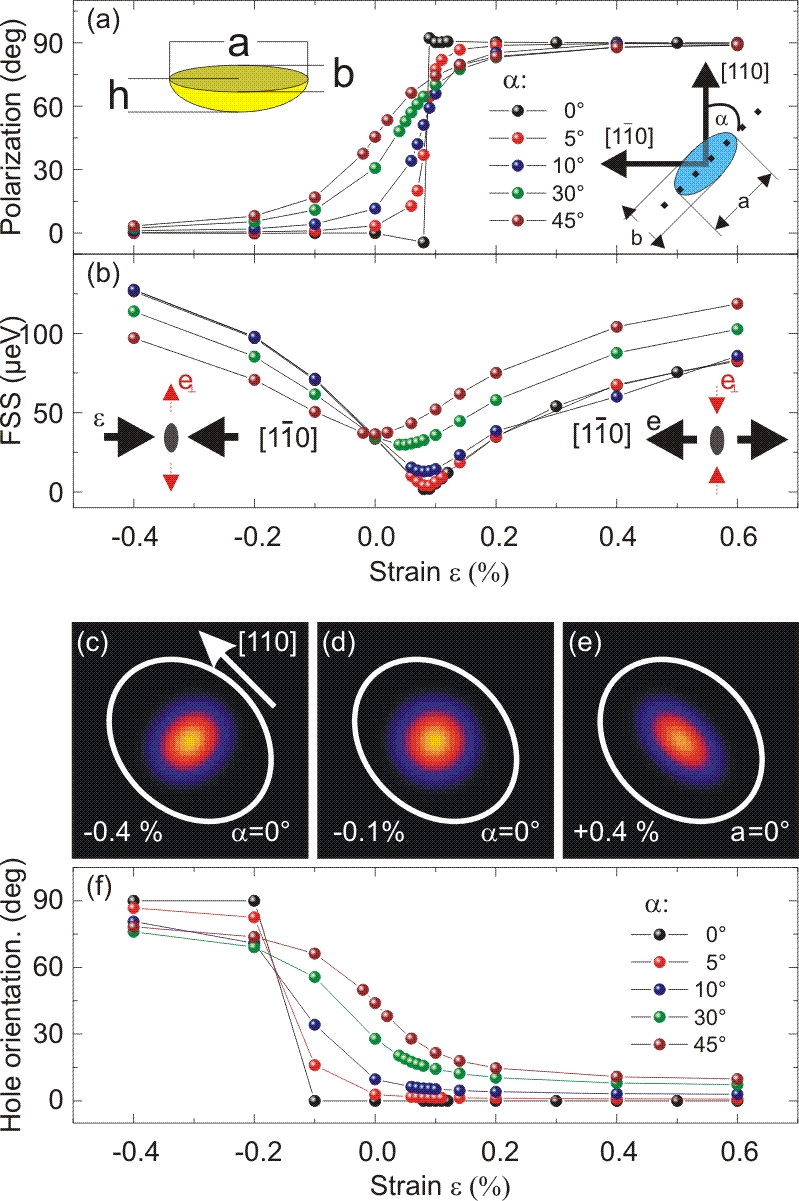}\caption{
(Color online) Theoretical study of the influence of anisotropic
biaxial strain on the light emitted by a excitons in model
GaAs/AlGaAs QDs. The left inset in (a) shows the shape of the
artificial structure. (a) Polarization of the high energy excitonic
component with respect to the x direction ([1$\overline{1}$0]) in
Fig.~\ref{fig:QDs}(b) for different values of $\alpha$ (i.e. angle
of the elongation direction with respect to the [110] crystal
direction,  see right inset). (b) Corresponding values of the FSS,
the left (right) inset disp lays the direction of the applied stress
for compressive (tensile) strains $\varepsilon$. (c)-(e) Density map
of the hole wave function for a QD with $\alpha$=0, for different
strain values. The white ellipse indicates the shape of the QD. (f)
Orientation of the hole wave function with respect to the [110]
direction vs strain. See text for more details.} \label{fig:Theory}
\end{figure}

To explain the physical origin of the dramatic changes in the
emission of the neutral excitons, we calculate the excitonic FSS of model dots
by combining the eight-band k$\cdot$p model and the configuration
interaction method following the approach described in
Refs.~\onlinecite{Stier99, Takagahara00, Plumhof10}. Strain
is introduced via the Pikus-Bir
Hamiltonian~\cite{Pikus74}. In the model we consider a
 semi-ellipsoidal
GaAs/AlGaAs QD with composition equal to the nominal one used in the
experiment. The main axis $\varepsilon$ of the anisotropic biaxial
strain [x axis in Fig.~\ref{fig:rawdata}(b)] coincides with the
[1$\overline{1}$0] GaAs crystal direction and we assume
$\varepsilon_{\bot}\approx -0.7 \times \varepsilon$ [see insets in
Fig.~\ref{fig:Theory}(b)]. The main axis of the QD forms an angle
$\alpha$ with respect to the [110] direction [see right inset in
Fig.~\ref{fig:Theory}(a)]~\cite{Supplementary}.

Figures~\ref{fig:Theory}(a) and (b) show, for different values of $\alpha$,
the calculated polarization angle and
 the FSS as a function of $\varepsilon$, respectively.
We begin with the ideal situation of a QD elongated along the [110]
crystal direction ($\alpha$=0). For zero strain, the QD has a FSS of
$33\ \mathrm{\mu
eV}$, which is in good agreement with typical observed values. When
the QD is stretched, both exciton transitions remain linearly
polarized and perpendicular to each other (not shown) and the FSS
varies in a wide range [see Fig.~\ref{fig:Theory}(b)]. For a strain
$\varepsilon$ of 0.086~\% the FSS reaches its minimum value below
0.4 $\mathrm{\mu eV}$. The polarization direction of the high energy
component, shown in Fig.~\ref{fig:Theory}(a), is perpendicular to
the elongation direction of the QD for strains below 0.086~\% and
abruptly changes by 90 degrees for higher strains.

 For increasing
$\alpha$, the polarization direction varies in a wider range of
strain values, i.e. it rotates smoothly as a function of strain as
shown in Fig.~\ref{fig:Theory}(a). Correspondingly, the minimum of
the FSS becomes increasingly broad and the reached minimum value
increases with increasing $\alpha$. We also note that at zero strain
the polarization angle is determined by the orientation of the dot.
As the strain increases, the structural orientation becomes less
important and the polarization angle is determined mostly by the
magnitude and direction of the strain, yielding similar values for
all dot orientations. This behavior is also observed in the
experiments: for the largest available strains, and away from the
FSS minimum, anisotropic biaxial stress allows us to orient the
polarization direction parallel/perpendicular to the strain
direction in a predictable way [see Figs.~\ref{fig:rawdata}(a)-(e)].
Finally we note that similar results are obtained when the direction
of the strain direction is changed and the QD is kept fixed instead
of rotating the QD shape with respect to the crystal direction as
 discussed here~\cite{Supplementary}.

Although it is difficult to ascribe the observed strain-induced
changes of the FSS to a single effect, we inspected the single
particle states and found that small strains produce relevant
changes on the hole wave function, while their effect on the
electron wave function is much weaker. In particular we find that:
(i) The proportion of the light hole band in the hole state is
substantially increased (e.g.\ from 0.6\,\% at {$\varepsilon$=0~\%}
to 11\,\% at {$\varepsilon$=0.2~\%} for $\alpha$=0); (ii) The hole
wave function changes in shape and orientation, as illustrated in
Fig.~\ref{fig:Theory}(c)-(e) for $\alpha$=0. In general, for finite
values of $\alpha$, the elongation direction of holes rotates by up
to 90$^{\circ}$, as shown in Fig.~\ref{fig:Theory}(f). At the same
time the electron wave function rotates by $<7^{\circ}$ (not shown).

These effects are a consequence of the non-diagonal terms in the
Pikus-Bir Hamiltonian, which enhance the mixing of the heavy hole
band with other bands (in particular, the light hole band) and
modify the effective mass causing its pronounced anisotropy along
the principal stress axes [110] and [1$\bar{1}$0]. The smooth
rotation of the hole wave function observed for finite values of
$\alpha$ [see Fig.~\ref{fig:Theory}(f)], which is analogous to the
 behavior observed for the polarization direction [see Fig.~\ref{fig:Theory}(a)], is
due to the joined influence of the structural anisotropy (elongation
of the QD) and the anisotropy of the effective mass, which tends to
elongate the wave functions along the [110] or [1$\bar{1}$0]
direction depending on the sign of the applied strain. We note that
the effects of the strain induced band edge shift and the created
piezoelectric potential on the FSS are negligible. Both effects are
comparably small (below 10~meV, for $\varepsilon$=0.2~\%) and do not
affect the lateral anisotropy of the wave functions.

In spite of the simplicity of the model QD shape, the presented
continuum model is able, for finite values of $\alpha$, to account
for all of the experimental observations apart from the fluctuations
in polarization angle observed in Figs.~\ref{fig:rawdata}(f) and
~\ref{fig:QDs}(a),(c). We tentatively ascribe the latter to
peculiarities of the shape and composition profiles of real QDs. We
thus conclude that atomistic models~\cite{Singh10,Bryant10} are not
strictly required to explore the main effects produced by stress on
the FSS and polarization angle of the excitonic states of QDs,
especially in view of the uncertainties in the structural parameters
of real QDs.

In conclusion, we have reported on the effects of anisotropic
biaxial stress on the emission of neutral excitons confined in
single semiconductor QDs. We have shown that relatively small
strains are sufficient to produce dramatic changes of the
polarization direction and of the energy splitting of the excitonic
exchange-split doublet. Qualitatively the same results are obtained
from three different kinds of QDs, consistent with a scenario
involving a strain-induced anticrossing of the bright excitonic
states~\cite{Singh10}. Based on a continuum model, which is able to
reproduce the main observed features, we ascribe the effects to
substantial changes of the hole states. The combination of
experimental and theoretical investigations of anisotropic stress
applied to QDs will allow to manipulate the excitonic
characteristics in terms of anisotropy and energy in a predictable
way.

We acknowledge V. Fomin, S. Kiravittaya, P. Atkinson, T. Zander, R.
Trotta, G. Bester, R. Singh and C. C. Bof Bufon for fruitful
discussions and technical support. V.~K. and P.~K. were supported by
Institutional research program MSM~0021622410 and the GACR grant
GA202/09/0676. This work was supported by the DFG (FOR730 and SFB
787).

%\bibliography{UnaxpaperBib-rev}

\end{document}